\documentclass[twocolumn,showpacs]{revtex4}

\usepackage[dvips]{graphicx}
\usepackage{color}

\begin{document}

\title{Teleportation-based number state manipulation
with number sum measurement}
\date{\today}
\author{Akira Kitagawa}
\email{kitagawa@nucleng.kyoto-u.ac.jp}
\author{Katsuji Yamamoto}
\email{yamamoto@nucleng.kyoto-u.ac.jp}
\affiliation{Department of Nuclear Engineering,
Kyoto University, Kyoto 606-8501, Japan}

\begin{abstract}
We examine various manipulations of photon number states
which can be implemented by teleportation technique
with number sum measurement.
The preparations of the Einstein-Podolsky-Rosen resources
as well as the number sum measurement resulting in projection
to certain Bell state may be done conditionally
with linear optical elements, i.e.,
beam splitters, phase shifters and zero-one-photon detectors.
Squeezed vacuum states are used as primary entanglement resource,
while single-photon sources are not required.
\end{abstract}


\maketitle 

\section{Introduction}
\label{sec:introduction}

Extensive research and development
have been done recently on quantum information technologies,
and teleportation is known to provide important tools
for quantum communication and information processing.
Its protocol was originally proposed by using a two-dimensional system,
qubit, such as the polarization of light
\cite{BBCJPW}.
Other protocols were also considered
to teleport the quadrature phase components
\cite{Braunstein,Milburn}
and the multi-dimensional states such as photon number states
\cite{BBCJPW,Milburn,Cochrane,Cochrane-Milburn}.
Among these media for quantum teleportation,
the photon number Fock space may be considered
more promising in the view point that it provides higher dimensional states
such as qutrits to carry more quantum information than qubits.

In order to perform faithfully the number state teleportation
in some finite dimensional subspace of Fock space,
maximally entangled Einstein-Podolsky-Rosen (EPR) resource is required.
Then, we have proposed recently a practical method
to generate such EPR states
\cite{KY-2002}.
That is, more entangled states are obtained
from a pair of squeezed vacuum states via swapping
by performing the number-phase Bell measurement.
In particular, the resultant states are maximally entangled
by adjusting the two squeezing parameters to the same value.
This may be viewed as entanglement concentration
\cite{concentration}.
The number state teleportation can be performed
by using these number-phase Bell states.
It is noticed in this teleportation
that some manipulations are made
such as shift and truncation of photon number states
and also scaling of amplitudes by the ratio of two squeezing parameters.
In this way, teleportation protocols may be considered
at the same time as those to manipulate input states
by projectictive measurements with EPR resources
(irrespective of teleportation fidelity).
In fact, quantum scissors protocol is investigated
for number state truncation by projective measurement
\cite{scissors1,scissors2,scissors2a}.
Therefore, we may expect that various manipulations
of number states are implemented via teleportation
without knowing the input states.
This method can also be used to transfer
to the output states the quantum information encoded in the EPR resources.

In this article, we examine various manipulations of number states
based on teleportation using projective measurement
of Bell state with certain number sum.
For this purpose, EPR resources with number difference 0
and those with number sum $ N $ are found to be useful.
These essential ingredients for teleportation,
EPR resources and projective measurements,
will be realized experimentally with linear optics,
specifically beam splitters, phase shifters
and zero-one-photon detectors.
Squeezed vacuum states are used as primary entanglement resource,
while single-photon sources are not required.
This may be an interesting point in the present scheme.

In Sec. \ref{sec:EPR} we consider the desired EPR resources.
Then, in Sec. \ref{sec:measurement} we describe
how the projective measurements to the number sum Bell states are realized
with linear optics.
In Sec. \ref{sec:t-m} some basic manipulations are obtained
via number state teleportations with these EPR resources
and the number sum measurement, and then by combining these tools
various manipulations are constructed.
Sec. \ref{sec:summary} is devoted to summary.

\section{EPR resource}
\label{sec:EPR}

We begin with examining the EPR resource
which may be shared by Bob and Alice
for number state teleportation.
We present in the following two kinds of EPR resources
with certain number difference and sum, respectively,
and consider their characteristic properties.

\subsection{Number difference resource}

An EPR state with number difference 0 is given generally
with amplitude distribution $ {\bf d} = ( d_1 , d_2 , \ldots ) $ by
\begin{equation}
| {\rm EPR} \rangle_{12}
= | 0 , {\bf d} \rangle^{(-)}_{12}
= \sum_{k=0}^\infty d_k | k \rangle_1 | k \rangle_2 .
\label{EPR-}
\end{equation}
Among such states, squeezed vacuum states generated
by parametric down conversion may be primarily interesting.
They are used widely as entanglement resource in various experiments
utilizing photons.  A two-mode squeezed vacuum state is described as
\begin{equation}
| \lambda \rangle_{12} = ( 1 - \lambda^2 )^{1/2}
\sum_{k=0}^\infty \lambda^k | k \rangle_1 | k \rangle_2 ,
\label{svs}
\end{equation}
where $ \lambda $ is the squeezing parameter.
This non-uniform number distribution with $ \lambda < 1 $
in Eq. (\ref{svs}) is attributed to the fact that the Fock space
is infinite dimensional while the energy should be finite
for the physical states.
Hence, the squeezed vacuum states are not maximally entangled.
We may also use a photon-subtracted state
which is generated from a squeezed vacuum state
\cite{OKW-2000-CRM-2002}:
\begin{equation}
| \lambda , - 1 \rangle_{12}
= {\sqrt{\frac{( 1 - \lambda^2 )^3}{1 + \lambda^2}}}
\sum_{k=0}^\infty ( k + 1 ) \lambda^k | k \rangle_1 | k \rangle_2 .
\label{svs-1}
\end{equation}
EPR states with number difference $ \Delta N \not= 0 $
may further be used,
with which the photon number shift is induced via teleportation.
They are, however, not considered in the following,
since the number shift of $ \Delta N = {\tilde N} - N $
is also realized by taking number sum $ N $ EPR states
and number sum $ {\tilde N} $ measurement, as seen in Sec. \ref{sec:t-m}.

\subsection{Number sum resource}

We next consider EPR states with number sum $ N $.
An $ N $-photon entangled state is given generally by
\begin{equation}
| {\rm EPR} \rangle_{12}
= | N , {\bf d} \rangle^{(+)}_{12}
= \sum_{k=0}^N d_k | N - k \rangle_1 | k \rangle_2 .
\label{EPR+}
\end{equation}
Specifically, we consider the Bell states
of number sum and phase difference
\cite{KY-2002,L-S,P-B},
\begin{equation}
| N , m \rangle_{12}
= \sum_{k=0}^N \frac{[ ( \omega_{N+1}^* )^m ]^k}{\sqrt{N+1}}
| N - k \rangle_1 | k \rangle_2 ,
\label{Bell-np}
\end{equation}
where $ m = 0 , 1 , \ldots , N $ $ {\rm mod} \ N + 1 $
($ | N , - 1 \rangle_{12} \equiv | N , N \rangle_{12} $, etc),
and the $ ( N+1 ) $-root is given by
\begin{equation}
\omega_{N+1} \equiv \exp \left[ i 2 \pi /(N+1) \right] , \
( \omega_{N+1} )^{N+1} = 1 .
\end{equation}
These Bell states with number sum $ N $
are also the eigenstates of the suitably defined Hermitian operator
of phase difference, and the eigenvalues are given by
\begin{equation}
\phi^{(N)}_m = \frac{2 \pi}{N+1} m .
\end{equation}
Here an arbitrary nonzero reference phase may be included
in $ \phi^{(N)}_m $ by a phase shift of the mode 2
as $ a_2 \rightarrow {\rm e}^{i \phi} a_2 $ in Eq. (\ref{Bell-np}),
though it does not change the following investigations.

As proposed in Ref. \cite{KY-2002},
we can generate the number-phase Bell states
by performing the number-phase measurement
on a pair of squeezed vacuum states.
Two squeezed vacuum states, 1-3 system and 2-4 system,
are rearranged via swapping (1-3, 2-4) $ \rightarrow $ (1-2, 3-4) as
\begin{eqnarray}
| \lambda \rangle_{13} | \lambda^\prime \rangle_{24}
&=& ( 1 - \lambda^2 )^{1/2} ( 1 - \lambda^{\prime 2} )^{1/2}
\sum_{N=0}^\infty \frac{K ( \lambda, \lambda^\prime , N )}{\sqrt{N+1}}
\nonumber
\\
& \times & 
\sum_{m=0}^N | N , - m \rangle_{34} | N , m , r \rangle_{12} .
\label{1234-bell}
\end{eqnarray}
Here, the generalized Bell states are introduced as
\begin{equation}
| N , m , r \rangle_{12}
= \frac{\lambda^N {\sqrt{N+1}}}{K ( \lambda, \lambda^\prime , N )}
\sum_{k=0}^N \frac{[ ( \omega_{N+1}^* )^m ]^k}{\sqrt{N+1}}
r^k | N - k \rangle_1 | k \rangle_2
\end{equation}
with the ratio of the squeezing parameters
\begin{equation}
r = \lambda^\prime / \lambda
\end{equation}
and the normalization factor
\begin{equation}
K ( \lambda, \lambda^\prime , N )
= \left[ \frac{\lambda^{2(N+1)} - \lambda^{\prime 2(N+1)}}
{\lambda^2 - \lambda^{\prime 2}} \right]^{1/2} .
\end{equation}
Then, the generalized Bell state $ | N , m , r \rangle_{12} $ is obtained
by the number-phase measurement of $ | N , - m \rangle_{34} $:
\begin{equation}
| \lambda \rangle_{13} | \lambda^\prime \rangle_{24} \Rightarrow
| N , m , r \rangle_{12} .
\end{equation}
In particular, they are maximally entangled
by adjusting the squeezing parameters
as $ \lambda = \lambda^\prime $, i.e.,
\begin{equation}
| N , m , r \rangle_{12}
\stackrel{\lambda = \lambda^\prime}{\longrightarrow}
| N , m \rangle_{12} \equiv | N , m , 1 \rangle_{12} .
\end{equation}
The probability to obtain the specific Bell state
$ | N , m , r \rangle_{12} $ by measuring $ | N , - m \rangle_{34} $
is given from Eq. (\ref{1234-bell}) as
\begin{equation}
P ( N , \lambda , \lambda^\prime )
= \frac{( 1 - \lambda^2 ) ( 1 - \lambda^{\prime 2} ) \lambda^{2N}}{N+1}
\frac{1 - r^{2(N+1)}}{1 - r^2} .
\label{PBell-Nr}
\end{equation}
This probability is evaluated for the typical cases of $ r $ as
\begin{equation}
P ( N , \lambda , \lambda^\prime )
\approx \left\{ \begin{array}{ll}
{\displaystyle{\frac{( 1 - {\bar \lambda}^2 ) {\bar \lambda}^{2N}}{N + 1}
}} &
( r^2 \gg 1 , r^2 \ll 1 ) \\
{} & {} \\
( 1 - \lambda^2 )^2 \lambda^{2N} &
( r \approx 1 ) \end{array} \right. ,
\end{equation}
where $ {\bar \lambda} = {\rm max} [ \lambda , \lambda^\prime ] $.
It is then estimated for generic $ r $ with optimal value of
$ {\bar \lambda} \approx [ 1 - (N+1)^{-1} ]^{1/2} \approx 0.7 - 0.9 $
for $ 1 \leq N \leq 5 $ as
\begin{equation}
P ( N , \lambda , \lambda^\prime )
\sim \frac{p(N)}{{\rm e} ( N + 1 )^2} ,
\label{PBell-Nr-1}
\end{equation}
where $ [ 1 - (N+1)^{-1} ]^N \approx 1/{\rm e} $ for $ N \gg 1 $
is considered.
The success probability $ p (N) $
for the measurement of $ | N , - m \rangle_{34} $ is included henceforth,
as will be given in Eq. (\ref{pN}), since it may be done conditionally.

There are of course other maximally entangled $ N $-photon states
than the number-phase Bell states, e.g.,
$ ( | 2 \rangle_1 | 0 \rangle_2 + i | 1\rangle_1 | 1 \rangle_2
+ | 0 \rangle_1 | 1 \rangle_2 ) / {\sqrt 3} $ with $ N = 2 $.
Some general methods to obtain $ N $-photon entangled states
have been investigated by utilizing linear optical devices
\cite{LKCD-2002-KLD-2002,F-2002}.

\section{Number sum measurement}
\label{sec:measurement}

\begin{figure}[t]
\scalebox{.5}{\includegraphics*[0cm,0cm][15cm,16cm]{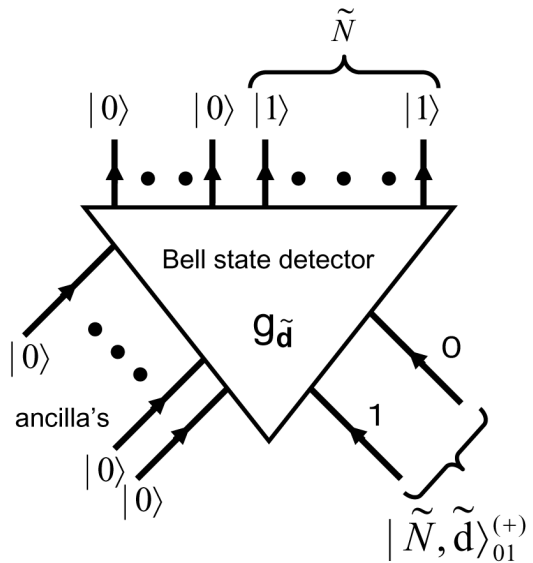}}
\caption{
A schematic diagram for the linear optical detector
of Bell state $ | {\tilde N} , {\tilde{\bf d}} \rangle^{(+)}_{01} $
with number sum $ {\tilde N} $.
The input two-mode state,
which may contain $ | {\tilde N} , {\tilde{\bf d}} \rangle^{(+)}_{01} $,
enters the detector together with the vacuum states
of several ancilla modes.
Only the Bell state $ | {\tilde N} , {\tilde{\bf d}} \rangle^{(+)}_{01} $
passing through this detector gives the output of the one photon states
in the $ {\tilde N} $ ports ($ 1^\prime $ to $ {\tilde N}^\prime $)
and the vacuum states in the other ports
with probability amplitude $ g_{\tilde{\bf d}} $.
}
\label{f-bell}
\end{figure}

We adopt the measurement resulting in projection to certain Bell state
with number sum $ {\tilde N} $,
\begin{equation}
| {\rm Bell} \rangle_{01}
= | {\tilde N} , {\tilde{\bf d}} \rangle^{(+)}_{01}
= \sum_{l=0}^{{\tilde N}}
{\tilde d}_l | {\tilde N} - l \rangle_0 | l \rangle_1 ,
\label{Bell+}
\end{equation}
where $ {\tilde N} $ may or may not be equal to $ N $
of the EPR resource $ | N , {\bf d} \rangle^{(+)}_{12} $.
Actually, as shown in Fig. \ref{f-bell},
the number sum measurement for the Bell state
$ | {\tilde N} , {\tilde{\bf d}} \rangle^{(+)}_{01} $
may be performed with some probability
by setting suitably linear optical elements, i.e.,
beam splitters, phase shifters and zero-one-photon detectors.
This method is based on the idea of photon chopping
\cite{Paul,Kok-Braunstein}.
Since the total number of photons is conserved
through this linear optical set,
the number sum $ {\tilde N} $ is measured
as the total count of photon detections.
The aim of the present analysis is to provide various manipulations
of photon number states with linear optics,
and hence we assume for clarity
that the zero-one-photon detectors can ideally discriminate
the photon numbers $ n $ = 0, 1 and more.
The estimate of efficiency in actual experiments
with imperfect detectors will be presented elsewhere
specifically for the manipulations of qubits and qutrits.
The single-photon detection is essential and quite challenging
for realizing quantum communication and processing with photon number states.
It is indeed encouraging that
some significant developments and new proposals have been made
for single-photon detection to achieve the quantum efficiency
close to unity \cite{SPD1,SPD2}.

The projection to the number sum Bell state
$ | {\tilde N} , {\tilde{\bf d}} \rangle^{(+)}_{01} $ in Eq. (\ref{Bell+})
will be done as follows
by attaching some number of {\it ancilla} modes
which are initially in the vacuum states $ | 0 \rangle_{\mbox{a's}} $.
By suitably setting the optical elements
and choosing their parameters, the desired Bell state may be transformed as
\begin{eqnarray}
| {\tilde N} , {\tilde{\bf d}} \rangle^{(+)}_{01}
| 0 \rangle_{\mbox{a's}}
& = & g_{\tilde{\bf d}} | 1 \rangle_{1^\prime}
\cdots | 1 \rangle_{{\tilde N}^\prime}
| 0 \rangle_{({\tilde N}+1)^\prime} | 0 \rangle_{({\tilde N}+2)^\prime}
\cdots
\nonumber \\
& + & \ldots
\end{eqnarray}
with nonzero coefficient $ g_{\tilde{\bf d}} $,
while any orthogonal state
$ | {\tilde N}_\bot , {\tilde{\bf d}}_\bot \rangle^{(+)}_{01} $
does not contain the first component,
where ``$ \ldots $" in the second line represents
the output states with other photon number distributions.
Then, by registering one photon in the output ports
$ 1^\prime $ to $ {\tilde N}^\prime $ and no photon in the others,
we can measure the Bell state
$ | {\tilde N} , {\tilde{\bf d}} \rangle^{(+)}_{01} $
conditionally with success probability
\begin{equation}
p ( {\tilde N} , {\tilde{\bf d}} ) = | g_{\tilde{\bf d}} |^2 .
\end{equation}

It should be mentioned that the number sum is not enough
to specify the Bell state in Eq. (\ref{Bell+})
with the amplitude distribution $ {\tilde{\bf d}} $.
Then, the joint measurement of number sum and phase difference
is usually adopted in number state teleportation
\cite{Milburn,Cochrane,Cochrane-Milburn,KY-2002},
which results in the projection to the Bell state
\begin{equation}
| {\tilde N} , \phi \rangle_{01}
= \sum_{l=0}^{\tilde N} \frac{({\rm e}^{- i \phi})^k}{\sqrt{{\tilde N}+1}}
| {\tilde N} - l \rangle_0 | l \rangle_1 .
\end{equation}
This state becomes $ | {\tilde N} , {\tilde m} \rangle_{01} $
with the discrete eigenvalue of phase difference
$ \phi = \phi^{({\tilde N})}_{\tilde m} $.
We may eliminate the phase factor $ ({\rm e}^{- i \phi})^k $
with a phase shifter on the mode 1 making the transformation
$ a_1 \rightarrow {\rm e}^{i \phi} a_1 $.
Hence, in the following we adopt specifically the projective measurement
of the $ {\tilde N} $-photon Bell state
with phase difference $ \phi = 0 $,
\begin{equation}
| {\tilde N} , 0 \rangle_{01}
= \frac{| {\tilde N} \rangle_0 | 0 \rangle_1
+ | {\tilde N} - 1 \rangle_0 | 1 \rangle_1
+ \ldots + | 0 \rangle_0 | {\tilde N} \rangle_1}{\sqrt{{\tilde N}+1}}
\end{equation}
with
\begin{equation}
{\tilde{\bf d}} = ( 1 , 1 , \ldots , 1 ) / {\sqrt{{\tilde N}+1}} .
\end{equation}

The projective measurement of $ | {\tilde N} , 0 \rangle_{01} $
may be done practically with linear optics.
The measurement of $ | 0 , 0 \rangle_{01} $ with $ {\tilde N} = 0 $
and $ | 1 , 0 \rangle_{01} $, $ | 1 , 1 \rangle_{01} $
with $ {\tilde N} = 1 $ is made at the same time
through a 50/50 beam splitter
with photon countings on the two output ports, as is well known.
For the case of $ | 2 , 0 \rangle_{01} $ with $ {\tilde N} = 2 $,
we have recently proposed a conditional method,
which would be extended to the cases of $ {\tilde N} \geq 3 $.
The number-phase Bell bases $ | {\tilde N}, {\tilde m} \rangle_{01} $
to be measured are explicitly given
for $ {\tilde N} = 0 $ with $ \omega_1 = 1 $,
$ {\tilde N} = 1 $ with $ \omega_2 = - 1 $
and $ {\tilde N} = 2 $ with $ \omega_3 \equiv \omega $, $ \omega^3 = 1 $
(the indices $ 0 $ and $ 1 $ denoting the modes omitted) as
\begin{eqnarray}
{\tilde N} = 0 :
&&
| 0 , 0 \rangle = | 0 \rangle | 0 \rangle ,
\\
{\tilde N} = 1 : &&
| 1 , 0 \rangle = ( | 1 \rangle | 0 \rangle
+ | 0 \rangle | 1 \rangle ) / {\sqrt 2} ,
\nonumber \\
&&
| 1 , 1 \rangle = ( | 1 \rangle | 0 \rangle
- | 0 \rangle | 1 \rangle ) / {\sqrt 2} ,
\\
{\tilde N} = 2 :
&&
| 2 , 0 \rangle = ( | 2 \rangle | 0 \rangle
+ | 1 \rangle | 1 \rangle
+ | 0 \rangle | 2 \rangle ) / {\sqrt 3} ,
\nonumber \\
&&
| 2 , 1 \rangle = ( | 2 \rangle | 0 \rangle
+ \omega^* | 1 \rangle | 1 \rangle
+ \omega | 0 \rangle | 2 \rangle ) / {\sqrt 3} ,
\nonumber \\
&&
| 2 , 2 \rangle = ( | 2 \rangle | 0 \rangle
+ \omega | 1 \rangle | 1 \rangle
+ \omega^* | 0 \rangle | 2 \rangle ) / {\sqrt 3} . \ \ \
\end{eqnarray}
We describe in the following the projective measurement
of these Bell states.

\begin{figure}[t]
\scalebox{.5}{\includegraphics*[0cm,0cm][15cm,16cm]{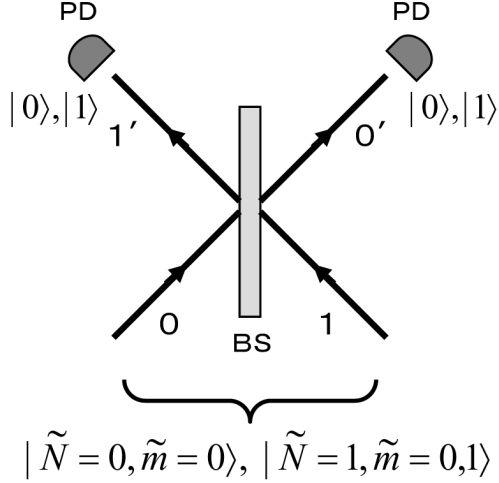}}
\caption{
A schematic diagram for the measurement of the Bell states
$ | {\tilde N} = 0 , {\tilde m} = 0 \rangle $
and $ | {\tilde N} = 1 , {\tilde m} = 0 , 1 \rangle $,
where ``BS" and ``PD" represent the beam splitter and photon detector,
respectively.
These Bell states with number sum $ {\tilde N} = 0 , 1 $
can be determined completely with the 50/50 beam splitter
from the result of the two photon detectors,
when they both detect the vacuum or one photon states.
}
\label{f-n01}
\end{figure}

A linear optical operation on the two modes $ i $ and $ j $
is made by a combination of beam splitter and phase shifter,
which is represented by a U(2) transformation as
\begin{equation}
\left( \begin{array}{c} a_i^\dagger \\ a_j^\dagger \end{array} \right)
= U_{(ij)} \left( \begin{array}{c} a_{i^\prime}^\dagger \\
                        a_{j^\prime}^\dagger \end{array} \right) ,
\label{U(2)}
\end{equation}
where
\begin{equation}
U_{(ij)} = \left( \begin{array}{cc}
c & -s \eta \\ s \xi & c \eta \xi \end{array} \right)_{(ij)}
\label{U(ij)}
\end{equation}
with $ c^2 + s^2 = 1 $ ($ c , s \geq 0 $) and $ | \eta | = | \xi | = 1 $.
Generally, it should be noted that any unitary transformation of
$ {\rm U} (N) $ among $ N $ photon modes can be realized in this way
\cite{Reck-Zeilinger}.
The number-phase Bell measurement can be done
for the cases of $ {\tilde N} = 0 , 1 $
by using a 50/50 beam splitter with
\begin{equation}
( c , s , \xi , \eta )_{(01)}
= ( 1/{\sqrt 2} , 1/{\sqrt 2} , 1 , 1 ) ,
\end{equation}
as shown in Fig. \ref{f-n01}.
The Bell states are transformed through the beam splitter as
\begin{eqnarray}
{\tilde N} = 0
& : &
| 0 , 0 \rangle_{01} = | 0 \rangle_{0^\prime} | 0 \rangle_{1^\prime} ,
\\
{\tilde N} = 1
& : &
| 1 , 0 \rangle_{01} = | 1 \rangle_{0^\prime} | 0 \rangle_{1^\prime} ,
\nonumber \\
& {} &
| 1 , 1 \rangle_{01} = - | 0 \rangle_{0^\prime} | 1 \rangle_{1^\prime} .
\end{eqnarray}
Then, if no photons are registered by the photon detectors,
the state $ | 0 , 0 \rangle_{01} $ is measured.
On the other hand, if we detect one photon in the mode $ 0^\prime $
($ 1^\prime $) and none in the mode $ 1^\prime $ ($ 0^\prime $),
we measure the state $ | 1 , 0 \rangle_{01} $ ($ | 1, 1 \rangle_{01} $).
The other results of the photon detectors indicate
the states with number sum more than 1.
In this special case with $ {\tilde N} = 0 , 1 $,
the Bell measurement can be done completely
with success probability
\begin{equation}
p ( {\tilde N} = 0 , 1 ) = 1 .
\end{equation}

\begin{figure}[t]
\scalebox{.5}{\includegraphics*[0cm,0cm][15cm,16cm]{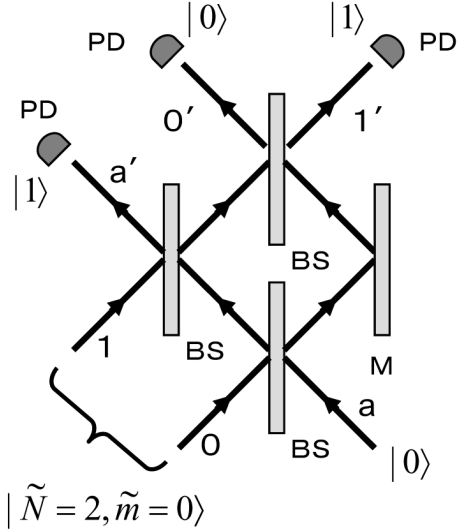}}
\caption{
A schematic diagram for the measurement of the Bell state
$ | {\tilde N} = 2 , {\tilde m} = 0 \rangle $,
where ``BS", ``PD" and ``M" represent
the beam splitter combined with phase shifter (if necessary),
photon detector and mirror, respectively.
By adjusting suitably the parameters of beam splitters,
as given in the text, the Bell state
$ | {\tilde N} = 2 , {\tilde m} = 0 \rangle $ to be measured
is identified conditionally
when the specific output state
$ | 0 \rangle_{0^\prime} | 1 \rangle_{1^\prime}
| 1 \rangle_{{\rm a}^\prime} $ is obtained by the three photon detectors.
}
\label{f-n2}
\end{figure}

The measurement of Bell state $ | 2 , 0 \rangle_{01} $
with $ {\tilde N} = 2 $ can be done conditionally,
as shown in Fig. \ref{f-n2}, introducing one ancilla mode ``$ {\rm a} $".
A series of unitary transformations such as given in Eq. (\ref{U(2)})
are made among these three modes through this optical set
\cite{Reck-Zeilinger}:
\begin{equation}
U = U_{(0{\rm a})} U_{(1{\rm a})} U_{(01)} ,
\end{equation}
where these unitary matrices are $ 3 \times 3 $ ones
by extending obviously Eq. (\ref{U(2)}).
The number-phase Bell states of $ {\tilde N} = 2 $
with the ancilla mode attached are given by
\begin{eqnarray}
| 2 , {\tilde m} \rangle_{01} | 0 \rangle_{\rm a}
& = & \sum_{i,j} C^{({\tilde m})}_{ij}
a_i^\dagger a_j^\dagger
| 0 \rangle_0 | 0 \rangle_1 | 0 \rangle_{{\rm a}}
\nonumber \\
& = & \sum_{i^\prime , j^\prime}
C^{({\tilde m}) \prime}_{i^\prime j^\prime}
a_{i^\prime}^\dagger a_{j^\prime}^\dagger
| 0 \rangle_{0^\prime} | 0 \rangle_{1^\prime} | 0 \rangle_{{\rm a}^\prime} .
\end{eqnarray}
These states are transformed with $ U $ as
\begin{equation}
C^{({\tilde m})} \rightarrow
C^{({\tilde m}) \prime} = U^{\rm T} C^{({\tilde m})} U ,
\end{equation}
and particularly
\begin{eqnarray}
| 2 , 0 \rangle_{01} | 0 \rangle_{\rm a}
& = &
g_0 | 0 \rangle_{0^\prime} | 1 \rangle_{1^\prime}
| 1 \rangle_{{\rm a}^\prime} + \ldots ,
\\
| 2 , 1 \rangle_{01} | 0 \rangle_{\rm a}
& = &
g_1 | 0 \rangle_{0^\prime} | 1 \rangle_{1^\prime}
| 1 \rangle_{{\rm a}^\prime} + \ldots ,
\\
| 2 , 2 \rangle_{01} | 0 \rangle_{\rm a}
& = &
g_2 | 0 \rangle_{0^\prime} | 1 \rangle_{1^\prime}
| 1 \rangle_{{\rm a}^\prime} + \ldots ,
\end{eqnarray}
where
$ g_{\tilde m} \equiv C^{({\tilde m}) \prime}_{1^\prime {\rm a}^\prime} $,
and ``$ \ldots $" represents the other orthogonal states.
It is here possible to adjust the parameters
of the linear operation $ U $ so as to give
\begin{equation}
g_0 \not= 0 , \ g_1 = 0 , \ g_2 = 0 .
\end{equation}
The relevant parameters are taken for example as
\begin{eqnarray}
( c , s , \eta , \xi )_{(0{\rm a})}
&=& \left( \frac{1}{\sqrt 2} , \frac{1}{\sqrt 2} , 1 , 1 \right) ,
\\
( c , s , \eta , \xi )_{(1{\rm a})}
&=& \left( {\sqrt{\frac{2}{3}}} , \frac{1}{\sqrt 3} , 1 ,
\frac{1 + i}{\sqrt 2} \right) ,
\\
( c , s , \eta , \xi )_{(01)}
&=& \left( {\sqrt{\frac{3}{8}}} , - {\sqrt{\frac{5}{8}}} ,
1 , \frac{3 + i}{\sqrt{10}} \right) .
\end{eqnarray}
Then, the state $ | 2 , 0 \rangle_{01} | 0 \rangle_{\rm a} $
only has the component of $ | 0 \rangle_{0^\prime} | 1 \rangle_{1^\prime}
| 1 \rangle_{{\rm a}^\prime} $,
while the states $ | 2, 1 \rangle_{01} | 0 \rangle_{\rm a} $
and $ | 2, 2 \rangle_{01} | 0 \rangle_{\rm a} $ do not.
This means that if we observe the state
$ | 0 \rangle_{0^\prime} | 1 \rangle_{1^\prime}
| 1 \rangle_{{\rm a}^\prime} $ with $ {\tilde N} = 2 $
by the photon detectors,
we measure the desired state $ | 2 , 0 \rangle_{01} $.
The probability to detect
$ | 0 \rangle_{0^\prime} | 1 \rangle_{1^\prime}
| 1 \rangle_{{\rm a}^\prime} $
as the conditional measurement of $ | 2 , 0 \rangle_{01} $
is determined with the above parameters as
\begin{equation}
p ( {\tilde N} = 2 ) = | g_0 |^2 = 3/8 .
\label{pN}
\end{equation}
Similarly, we can measure the Bell states $ | 2, \pm 1 \rangle_{01} $.
It is even desired that
this method with linear optical operations
for the conditional Bell measurement may be extended
to the cases of $ {\tilde N} \geq 3 $.

\section{Teleportation and manipulation}
\label{sec:t-m}

We now investigate the number state teleportation
with number sum measurement.
In general, the input state (mode 0) is prepared as
\begin{equation}
| \psi_{\rm in} \rangle_0 = \sum_{n=0}^\infty c^{\rm in}_n | n \rangle_0 .
\end{equation}
Alice and Bob share some entanglement resource $ | {\rm EPR} \rangle_{12} $,
as presented in Sec. \ref{sec:EPR}.
Then, Alice performs the number sum measurement on the modes 0 and 1,
as described in Sec. \ref{sec:measurement}.
Then, after Alice succeeds in the measurement
of $ | {\tilde N} , 0 \rangle_{01} $ with some probability,
Bob obtains the output state in the mode 2 as
\begin{eqnarray}
| \psi_{\rm out} \rangle_2
&=& {}_{10} \langle {\tilde N} , 0 | | \psi_{\rm in} \rangle_0
| {\rm EPR} \rangle_{12}
\nonumber
\\
&=& \sum_{n=0}^\infty
c^{\rm out}_n | n \rangle_2 .
\end{eqnarray}
We here do not suppose the unitary transformation by Bob for simplicity.
It is interesting that experimental results have been reported recently
for the teleportation of vacuum-one-photon qubit
based on this Bell measurement with $ {\tilde N} = 1 $
\cite{LSPM,scissors2a}.

\begin{table*}[t]
\caption{
\label{tab:basic}
Some basic manipulations are listed together with the EPR resources.
The projective measurement is made for the Bell state
$ | {\tilde N} , 0 \rangle $ with number sum $ {\tilde N} $
and phase difference 0.
The overall factors in the resultant output states are omitted
for simplicity.
(a) Reversal + scaling rearranges the number state amplitudes
in reverse order together with scaling by the squeezing parameter
$ \lambda $.
(b) Reversal + derivative rearranges the number state amplitudes
in reverse order together with differentiation with respect to $ \lambda $.
(c) Number shift makes addition or subtraction of $ | \Delta N | $ photons.
(d) Scaling makes magnification of the number state amplitudes
by the powers of $ r $ (the ratio of the squeezing parameters).
}
\begin{ruledtabular}
\begin{tabular}{ccccc}
{} & (a) Reversal + scaling & (b) Reversal + derivative
   & (c) Number shift & (d) Scaling
\\
\hline
$ | {\rm EPR} \rangle $
 & $ | \lambda \rangle $
 & $ | \lambda , - 1 \rangle $
 & $ | N , 0 , r = 1 \rangle $
 & $ | N = {\tilde N} , 0 , r \rangle $
\\
Output $ c^{\rm out}_n $
 & $ \lambda^n c^{\rm in}_{{\tilde N} - n} $
 & $ ( n + 1 ) \lambda^n c^{\rm in}_{{\tilde N} - n} $
 & $ c^{\rm in}_{n + \Delta N} $ ($ \Delta N = {\tilde N} - N $)
 & $ r^n c^{\rm in}_n $ ($ r = \lambda^\prime / \lambda $)
\\
Photon number range $ n $
$ \begin{array}{l} {\rm input} \\
                   {\rm output} \end{array} $
 & $ \begin{array}{l} {\tilde N} , \ldots , 0 \\
                      0 , \ldots , {\tilde N} \end{array} $
 & $ \begin{array}{l} {\tilde N} , \ldots , 0 \\
                      0 , \ldots , {\tilde N} \end{array} $
 & $ \begin{array}{l} {\rm max}[ 0 , \Delta N ] , \ldots , {\tilde N} \\
                      {\rm max}[ 0 , - \Delta N ] , \ldots , N \end{array} $
 & $ \begin{array}{l} 0 , \ldots , N \\
                      0 , \ldots , N \end{array} $
\end{tabular}
\end{ruledtabular}
\end{table*}

\subsection{Number difference resource}

The output state using an EPR resource
$ | 0 , {\bf d} \rangle^{(-)}_{12} $ with number difference 0
is given as
\begin{equation}
| \psi_{\rm out} \rangle_2
= \frac{1}{\sqrt{{\tilde N} + 1}} \sum_{k=0}^{{\tilde N}}
d_k c^{\rm in}_{{\tilde N} - k} | k \rangle_2
\end{equation}
with ($ k \rightarrow n $)
\begin{equation}
c^{\rm out}_n = d_n c^{\rm in}_{{\tilde N} - n} / {\sqrt{{\tilde N} + 1}} \
( 0 \leq n \leq {\tilde N} ) .
\end{equation}
Hence, this teleportation realizes
the reversal of number distribution
as $ ( 0 , 1 , \ldots , {\tilde N} )
\rightarrow ( {\tilde N} , \ldots , 1 , 0 ) $,
and the information of $ {\bf d} $ in the EPR resource
is encoded in the output state.

In particular, using the squeezed vacuum state $ | \lambda \rangle $
we obtain a manipulation,
\begin{equation}
c^{\rm in}_n
\rightarrow \lambda^n c^{\rm in}_{{\tilde N} - n} \
( 0 \leq n \leq {\tilde N} ) ,
\end{equation}
where the overall factor independent of $ n $ is omitted for simplicity.
The probability
$ P = {}_2 \langle \psi_{\rm out} | \psi_{\rm out} \rangle_2 $
to obtain this output state is calculated
depending on the relevant parameters by
\begin{equation}
P ( {\tilde N} , \lambda , {\bf c}^{\rm in} )
= \frac{( 1 - \lambda^2 )}{{\tilde N} + 1} \sum_{n=0}^{{\tilde N}}
\lambda^{2n} | c^{\rm in}_{{\tilde N} - n} |^2 .
\label{P0-0}
\end{equation}
Here the success probability $ p ({\tilde N}) $
for the conditional measurement of $ | {\tilde N} , 0 \rangle_{12} $,
as given in Eq. (\ref{pN}), is omitted for simplicity.

By using the photon-subtracted state $ | \lambda , - 1 \rangle $,
we also obtain a manipulation,
\begin{equation}
c^{\rm in}_n
\rightarrow ( n + 1 ) \lambda^n c^{\rm in}_{{\tilde N} - n} \
( 0 \leq n \leq {\tilde N} ) ,
\end{equation}
corresponding to the derivative
$ d \lambda^{n+1} / d \lambda = ( n + 1 ) \lambda^n $.
The success probability is calculated as
\begin{eqnarray}
P ( {\tilde N} , \lambda , - 1 , {\bf c}^{\rm in} )
& = & \frac{( 1 - \lambda^2 )^3}{( {\tilde N} + 1 )( 1 + \lambda^2 )}
\nonumber \\
& \times & \sum_{n=0}^{{\tilde N}}
( n + 1 )^2 \lambda^{2n} | c^{\rm in}_{{\tilde N} - n} |^2 .
\label{P0-1}
\end{eqnarray}

\subsection{Number sum resource}

The output state using an EPR resource
$ | N , {\bf d} \rangle^{(+)}_{12} $ with number sum $ N $
is given as
\begin{equation}
| \psi_{\rm out} \rangle_2
= \frac{1}{\sqrt{{\tilde N} + 1}} \sum_{k=n_0}^N
d_k c^{\rm in}_{k + \Delta N} | k \rangle_2
\end{equation}
with
\begin{equation}
c^{\rm out}_n = d_n c^{\rm in}_{n + \Delta N} / {\sqrt{{\tilde N} + 1}} \
( n_0 \leq n \leq N ) ,
\end{equation}
where
\begin{eqnarray}
\Delta N & = & {\tilde N} - N ,
\\
n_0 & = & {\rm max}[ 0 , - \Delta N ] .
\end{eqnarray}
Hence, this teleportation realizes the number shift of $ \Delta N $ as
\begin{equation}
| n + \Delta N \rangle \rightarrow | n \rangle .
\end{equation}
Due to the conditions $ 0 \leq k \leq N $
and $ 0 \leq l \leq {\tilde N} $ with $ l = N - k $,
the sum is taken over the photon number $ n $ ($ k \rightarrow n $)
in the output state as $ n_0 \leq n \leq N $, i.e.,
\begin{equation}
n ({\rm out}) = \left\{ \begin{array}{ll}
0 , \ldots , N & ( {\tilde N} \geq N )
\\ | \Delta N | , \ldots , N & ( {\tilde N} < N ) \end{array} \right. .
\label{n-range}
\end{equation}
Here it should be noted that for the case of $ {\tilde N} < N $
the number state is truncated below $ | \Delta N | $ as well as above $ N $.

By using specifically the generalized number-phase Bell state
$ | N , 0 , r \rangle_{12} $, we obtain a manipulation,
\begin{equation}
c^{\rm in}_n
\rightarrow r^n c^{\rm in}_{n + \Delta N} \ ( n_0 \leq n \leq N )
\label{rDN}
\end{equation}
with the probability
\begin{equation}
P ( {\tilde N} , N , r , {\bf c}^{\rm in} )
= \frac{1 - r^2}{1 - r^{2(N+1)}}
\sum_{n=n_0}^N \frac{r^{2n} | c^{\rm in}_{n + \Delta N} |^2}
{{\tilde N}+1} .
\label{PNr}
\end{equation}
That is, the scaling by $ r $,
\begin{equation}
a^\dagger \rightarrow r a^\dagger ,
\end{equation}
is realized in addition to the number shift of $ \Delta N $.
The scaling factor $ r = \lambda^\prime / \lambda $
may be larger or smaller than 1
by adjusting the two squeezing parameters.
This manipulation (\ref{rDN}) may be decomposed
to the number shift of $ \Delta N $
with $ r = 1 $ ($ \lambda = \lambda^\prime $)
and the scaling by $ r $ with $ {\tilde N} = N $.

\subsection{Various manipulations}

We have obtained some basic manipulations of number states
via teleportation, where the projective measurement of the Bell state
$ | {\tilde N} , 0 \rangle $
with number sum $ {\tilde N} $ and phase difference 0 is adopted.
They are listed together with the EPR resources in Table \ref{tab:basic}:
(a) Reversal + scaling rearranges the number state amplitudes
in reverse order together with scaling by the squeezing parameter
$ \lambda $.
(b) Reversal + derivative rearranges the number state amplitudes
in reverse order together with differentiation with respect to $ \lambda $.
(c) Number shift makes addition or subtraction of $ | \Delta N | $ photons.
(d) Scaling makes magnification of the number state amplitudes
by the powers of $ r $ (the ratio of the squeezing parameters).
In these manipulations, the number states are truncated
by the photon numbers of the Bell state and the EPR resources.
The success probabilities of these manipulations
may be evaluated in Eqs. (\ref{P0-0}), (\ref{P0-1}) and (\ref{PNr})
typically for the state with uniform number distribution
up to $ {\tilde N} $,
i.e., $ | c^{\rm in}_n |^2 = 1/( {\tilde N} + 1 ) $
for $ n = 0 , 1 , \ldots {\tilde N} $
and $ c^{\rm in}_n = 0 $ for $ n \geq {\tilde N} + 1 $:
\begin{eqnarray}
P_{(a)} & = & \frac{p ({\tilde N})}{( {\tilde N} + 1 )^2}
( 1 - \lambda^{2({\tilde N}+1)} ) ,
\label{Pa}
\\
P_{(b)} & = & \frac{p ({\tilde N})}{( {\tilde N} + 1 )^2}
\frac{( 1 - \lambda^2 )^3}{( 1 + \lambda^2 )}
\sum_{n=0}^{{\tilde N}} ( n + 1 )^2 \lambda^{2n} ,
\label{Pb}
\\
P_{(c)} & = & \frac{p ({\tilde N})}{( {\tilde N} + 1 )^2}
\left( 1 - \frac{n_0}{N + 1} \right) ,
\label{Pc}
\\
P_{(d)} & = & \frac{p ({\tilde N})}{( {\tilde N} + 1 )^2} .
\label{Pd}
\end{eqnarray}
Here the success probability $ p ({\tilde N}) $
for the conditional measurement of $ | {\tilde N} , 0 \rangle_{12} $
is also included, as given in Eq. (\ref{pN}).

By combining these teleportation-based tools (a) -- (d),
we can perform various manipulations of number states,
as presented in the following.
The net success probabilities are roughly estimated
with $ P_{(a)} $ -- $ P_{(d)} $ in Eqs. (\ref{Pa}) -- (\ref{Pd}).
They may appear to be somewhat small
since the success probability $ P ( N , \lambda , \lambda^\prime ) $
as given in  Eq. (\ref{PBell-Nr-1}) is further multiplied
for the manipulations (c) and (d)
to prepare the EPR resource $ | N , 0 , r \rangle $.

The genuine reversal of number state can be realized
since the scaling associated with the manipulation (a)
can be cancelled by the manipulation (d)
with suitably adjusted squeezing parameters:
\begin{eqnarray}
&&
{\mbox{(d)}} [ r = \lambda^\prime / \lambda = 1 / \lambda^{\prime \prime} ]
\bullet {\mbox{(a)}} [ \lambda^{\prime \prime} ]
\nonumber \\
&& \Rightarrow
( c^{\rm in}_0 , \ldots , c^{\rm in}_{\tilde N} )
\rightarrow ( c^{\rm in}_{\tilde N} , \ldots , c^{\rm in}_0 ) .
\label{reversal}
\end{eqnarray}
The net success probability for the reversal is estimated
from Eqs. (\ref{Pa}) and (\ref{Pd}) together with Eq. (\ref{PBell-Nr-1}) as
\begin{equation}
P({\rm reversal})
\sim \frac{p ({\tilde N})^3}{{\rm e} ( {\tilde N} + 1 )^6} .
\end{equation}
Numerically, we have $ P({\rm reversal}) = 6 \times 10^{-3} $
for a qubit $ ( | 0 \rangle + 0.5 | 1 \rangle ) / {\sqrt{1 + 0.5}} $
with $ \lambda = 0.49 $, $ \lambda^\prime = 0.7 $,
$ \lambda^{\prime \prime} = 0.7 $ and $ p({\tilde N} = 1) = 1 $,
and $ P({\rm reversal}) = 2 \times 10^{-5} $
for a qutrit $ ( | 0 \rangle + 0.5 | 1 \rangle + 0.5^2 | 2 \rangle ) /
{\sqrt{1 + 0.5 + 0.5^2}} $
with $ \lambda = 0.7 $, $ \lambda^\prime = 0.49 $,
$ \lambda^{\prime \prime} = 0.7 $ and $ p({\tilde N} = 2) = 3/8 $.

On the other hand, by making the (a) reversal + scaling twice,
we obtain the (d) scaling as
\begin{equation}
{\mbox{(a)}} [ \lambda^\prime ] \bullet {\mbox{(a)}} [ \lambda ]
\Rightarrow {\mbox{(d)}} [ r = \lambda^\prime / \lambda ] ,
\label{d=aa}
\end{equation}
i.e.,
\begin{eqnarray}
c^{\rm in}_n
& \rightarrow & \lambda^n c^{\rm in}_{{\tilde N} - n}
\nonumber \\
& \rightarrow & \lambda^{\prime n} \left[
\lambda^{{\tilde N} - n} c^{\rm in}_{{\tilde N}-({\tilde N} - n)} \right]
= \lambda^{\tilde N} r^n c^{\rm in}_n .
\end{eqnarray}
This really corresponds to the fact that
two number-phase measurements are made
for the number state teleportation with the number-phase Bell states,
as described in Ref. \cite{KY-2002}.

We can also make some manipulations by using the (c) number shift
with truncation.  By choosing $ {\tilde N} = N $ ($ \Delta N = 0 $)
in the (c), i.e., $ r = 1 $ in the (d), we obtain the quantum scissors
\cite{scissors1,scissors2,scissors2a},
\begin{equation}
( c^{\rm in}_0 , \ldots , c^{\rm in}_N , c^{\rm in}_{N+1} , \ldots )
\rightarrow ( c^{\rm in}_0 , \ldots , c^{\rm in}_N , 0 , \ldots ) .
\end{equation}
The success probability is estimated with Eqs. (\ref{PBell-Nr-1})
and (\ref{Pc}) as
\begin{equation}
P({\rm scissors}; N )
\sim \frac{p (N)^2}
{{\rm e} ( N + 1 )^4} .
\end{equation}
Furthermore, by using the (c) number shift twice
with $ ( N , {\tilde N} ) = ( N_2 - N_1 , N_2 ) $
($ \Delta N = N_1 \geq 0 $)
and then $ ( N , {\tilde N} ) = ( N_2 , N_2 - N_1 ) $
($ \Delta N = - N_1 \leq 0 $),
we obtain another kind of scissors which cuts both sides
of the number state ($ n < N_1 $ and $ n > N_2 $) as
\begin{eqnarray}
&& ( c^{\rm in}_0 , \ldots ,
c^{\rm in}_{N_1} , \ldots , c^{\rm in}_{N_2} , \ldots )
\nonumber \\
&& \rightarrow
( c^{\rm in}_{N_1} , \ldots , c^{\rm in}_{N_2} , 0 , \ldots ) 
\nonumber \\
&& \rightarrow
( 0 , \ldots , 0 , c^{\rm in}_{N_1} , \ldots , c^{\rm in}_{N_2} ,
0 , \ldots ) ,
\end{eqnarray}
i.e.,
\begin{equation}
\sum_{n=0}^\infty c^{\rm in}_n | n \rangle
\rightarrow \sum_{n=N_1}^{N_2} c^{\rm in}_n | n \rangle ,
\label{scissors12}
\end{equation}
where the normalization factor in the output state is omitted.
The net success probability is estimated with Eqs. (\ref{PBell-Nr-1})
and (\ref{Pc}) as
\begin{equation}
P({\rm scissors}; N_1 , N_2 )
\sim \frac{p ( N_2 - N_1 )^2 p ( N_2 )^2}
{{\rm e}^2 ( N_2 - N_1 + 1 )^3 ( N_2 + 1 )^5} .
\label{Pscissors}
\end{equation}

By choosing particularly $ N_1 = N_2 = N \geq 0 $ in Eq. (\ref{scissors12}),
we obtain a quantum extractor as
\begin{equation}
\sum_{n=0}^\infty c^{\rm in}_n | n \rangle
\rightarrow | N \rangle \ ( | c^{\rm in}_N | \not= 0 ) ,
\label{extractor}
\end{equation}
transmitting only the specific number state $ | N \rangle $.
If $ c^{\rm in}_N = 0 $ incidentally,
the combined measurements with $ N_1 = N_2 = N $
as mentioned above to extract $ | N \rangle $ succeed with zero probability;
the state $ | N \rangle $ cannot be extracted
if the input state does not contain it.
Single-photon quantum-nondemolition detectors
with linear optics and projective measurements
have been considered recently
\cite{KLD-2002}.
This quantum extractor in fact acts
as $ N $-photon quantum-nondemolition detector
for $ N = 0 , 1 , 2 $, and so on.
The success probability is estimated in Eq. (\ref{Pscissors})
with $ N_1 = N_2 = N $ and $ p(0) = 1 $ as
\begin{equation}
P({\rm extractor}; N )
\sim \frac{p (N)^2}{{\rm e}^2 ( N + 1 )^5} .
\end{equation}

\begin{figure}[t]
\scalebox{.5}{\includegraphics*[0cm,0cm][15cm,16cm]{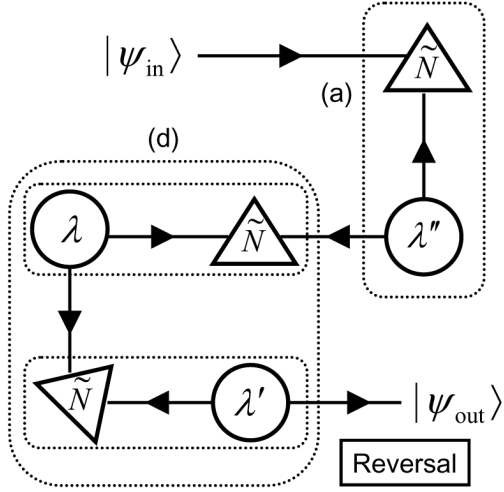}}
\caption{
A chain of the elements to realize the reversal of photon number states,
as described in Eq. (\ref{reversal}).
The circles and triangles represent the squeezed vacuum states
and Bell measurements, respectively.
The manipulations (a) and (d) are indicated with dotted boxes.
It is also seen that the (d) is composed of the two (a)'s,
as described in Eq. (\ref{d=aa}).
}
\label{f-reversal}
\end{figure}

The $ N $-photon source may be obtained more efficiently
in another way from the vacuum state $ | 0 \rangle $
with the EPR resource $ | N , 0 , 1 \rangle $
and the measurement of $ | 0 , 0 \rangle $ as
\begin{equation}
{\mbox{(c)}} [ N ; {\tilde N} = 0 ]
\Rightarrow | 0 \rangle \rightarrow | N \rangle .
\end{equation}
The success probability is estimated from Eqs. (\ref{PBell-Nr-1})
and (\ref{Pc}) with $ p ({\tilde N} = 0 ) = 1 $ as
\begin{equation}
P( | 0 \rangle \rightarrow | N \rangle )
\sim \frac{p(N)}{{\rm e} ( N + 1 )^3} .
\end{equation}
Numerically, we have
$ P( | 0 \rangle \rightarrow | 1 \rangle ) = 7 \times 10^{-2} $
with $ p ({\tilde N} = 1 ) = 1 $ and $ \lambda = 0.5 $
for the single-photon source,
and $ P( | 0 \rangle \rightarrow | 2 \rangle ) = 8 \times 10^{-3} $
with $ p ({\tilde N} = 2 ) = 3/8 $ and $ \lambda = 0.7 $
for the two-photon source.

We can even perform the differentiation of polynomials
by combining the reversal (\ref{reversal}),
the (b) reversal + derivative
and the (c) number shift with $ \Delta N = 1 $ ($ N = {\tilde N} - 1 $):
\begin{eqnarray}
c^{\rm in}_n
& \rightarrow & c^{\rm in}_{{\tilde N} - n} ( 0 \leq n \leq {\tilde N} )
\nonumber \\
& \rightarrow &
( n + 1 ) \lambda^n c^{\rm in}_{{\tilde N} - ( {\tilde N} - n )}
( 0 \leq n \leq {\tilde N} )
\nonumber \\
& \rightarrow & ( n + 1 ) \lambda^n c^{\rm in}_{n+1}
( 0 \leq n \leq N = {\tilde N} - 1 ) ,
\end{eqnarray}
i.e.,
\begin{equation}
f( \lambda ) = \sum_{n=0}^{\tilde N} c^{\rm in}_n \lambda^n
\rightarrow \frac{d f}{d \lambda}
= \sum_{n=0}^{{\tilde N} - 1} c^{\rm in}_{n+1} ( n + 1) \lambda^n .
\end{equation}

These teleportation-based manipulations as examined so far
are in fact built up by connecting the EPR resources and Bell measurements.
For example, shown in Fig. \ref{f-reversal} is
a chain of the elements to realize the reversal of photon number states,
as described in Eq. (\ref{reversal}).
The circles and triangles represent the squeezed vacuum states
and Bell measurements, respectively.
The manipulations (a) and (d) are indicated with dotted boxes.
It is also seen that the (d) is composed of the two (a)'s,
as described in Eq. (\ref{d=aa}).
The photon-subtracted EPR resource in Eq. (\ref{svs-1})
for the manipulation (b) is also obtained
in these diagrams by attaching a beam splitter and a single-photon detector
to each mode of a squeezed vacuum state
\cite{OKW-2000-CRM-2002}.
In this way, these manipulations form a semi-group
as chains of the EPR resources and Bell measurements.
The inverse, however, does not exist for any of them,
since the number state is truncated by the number sum $ {\tilde N} $
of the Bell measurement.

The number state manipulations may also be applied
to the multi-mode states.
It is interesting that the truncated but ideally squeezed vacuum state
is obtained by performing the (d) scaling of $ r = 1 / \lambda $
on the one mode of $ | \lambda \rangle $
so as to cancel the squeezing parameter $ \lambda < 1 $.
It achieves maximal entanglement with $ \lambda = 1 $
in Eq. (\ref{svs}) but finite dimensional,
as used for the teleportation in generic Hilbert spaces
\cite{BBCJPW}:
\begin{eqnarray}
| \lambda = 1 , N \rangle
& = & \frac{1}{\sqrt{N+1}} \sum_{n=0}^N | n \rangle | n \rangle
\nonumber \\
& = & \frac{| 0 \rangle | 0 \rangle + \ldots + | N \rangle | N \rangle}
{\sqrt{N+1}} .
\end{eqnarray}
The success probability is estimated from Eqs. (\ref{PBell-Nr})
and (\ref{PNr})
with $ r = \lambda^{\prime \prime} / \lambda^\prime = 1 / \lambda $
and $ c^{\rm in}_n = ( 1 - \lambda^2 )^{1/2} \lambda^n $:
\begin{equation}
P( | \lambda = 1 , N \rangle )
= \frac{p(N)^2}{(N+1)^2}
( 1 - \lambda^2 ) ( 1 - \lambda^{\prime \prime 2} )
( 1 - \lambda^{\prime 2} ) \lambda^{\prime 2N} .
\end{equation}
Numerically, we have
$ P( | \lambda = 1 , N = 1 \rangle ) = 1 \times 10^{-2} $
for $ ( | 0 \rangle | 0 \rangle + | 1 \rangle | 1 \rangle ) / {\sqrt 2} $
and $ P( | \lambda = 1 , N = 2 \rangle ) = 2 \times 10^{-4} $
for $ ( | 0 \rangle | 0 \rangle + | 1 \rangle | 1 \rangle
+ | 2 \rangle | 2 \rangle ) / {\sqrt 3} $, respectively,
with $ \lambda = 0.7 $, $ \lambda^{\prime} = 0.49 $,
$ \lambda^{\prime \prime} = 0.7 $and $ p(N=2) = 3/8 $.
More generally, by applying the scissors (\ref{scissors12})
we obtain
\begin{equation}
| \lambda = 1 , N_1 , N_2 \rangle
= \frac{| N_1 \rangle | N_1 \rangle + \ldots + | N_2 \rangle | N_2 \rangle}
{\sqrt{( N_2 - N_1 ) + 1}} .
\end{equation}
Furthermore, we can make the munipulations as
\begin{equation}
| \lambda \rangle \rightarrow | N \rangle | N \rangle
\rightarrow | N - \Delta N \rangle | N \rangle .
\end{equation}

It should here be mentioned that an interesting scheme has been proposed
recently for multi-mode operations on number states
which are implemented conditionally
with linear optical operations and zero-one-photon detections
\cite{CKW-2003}.
It will be relevant to the present investigation
for teleportation-based manipulations.
While we have used specific EPR resources so far,
we may prepare on demand the EPR resources as templates
for certain manipulations.
For example, if we prepare the $ N $-photon EPR resource
$ | N , {\bf d} \rangle^{(+)}_{12} $
with $ d_{N_1} = 0 $ for only one component of $ k = N_1 $ in $ {\bf d} $,
we obtain the quantum filter of $ | N_1 \rangle $ state.
In this way, we can make various manipulations
of number states via teleportation with number sum measurement.

\section{Summary}
\label{sec:summary}

In summary, we have investigated the number state manipulation
based on teleportation with number sum measurement.
In the present scheme, squeezed vacuum states
with number difference 0 are used as primary entanglement resource,
while single-photon sources are not required.
The projective measurement of the Bell state with certain number sum
and phase difference 0 is adopted specifically
as the number sum measurement,
which may be done with linear optical elements, i.e.,
beam splitters, phase shifters and zero-one-photon detectors.
The number-phase Bell states as $ N $-photon EPR resource
are really generated from a pair of squeezed vacuum states
by the number sum measurement.
Some basic manipulations are obtained
via teleportation with these EPR resources
and the number sum Bell measurement.
Then, by combining these basic tools
we can perform various manipulations of number states.
The desired generation of these EPR resources
as well as the number sum measurement will be realized
experimentally with linear optics.

\begin{acknowledgments}
The authors would like to thank M. Senami for valuable discussions.
\end{acknowledgments}

\end{document}